\documentclass[pra,aps,twocolumn,tightlines]{revtex4}
\usepackage{amsmath}
\usepackage{epsfig}

\begin{document}

\title{Generation of Kerr Frequency Combs in Resonators with Normal GVD}

\author{ A. B. Matsko, A. A. Savchenkov, and L. Maleki}

\affiliation{OEwaves Inc., 465 N. Halstead St. Ste. 140, Pasadena, CA 91107}

\begin{abstract}
We show via numerical simulation that Kerr frequency combs can be generated in a nonlinear resonator characterized with normal group velocity dispersion (GVD). We find the spectral shape of the comb and temporal envelope of the corresponding optical pulses formed in the resonator.
\end{abstract}

\maketitle

Kerr optical frequency combs have recently attracted considerable attention because of their rich physics and their potential for multiple applications in science and technology \cite{kippenberg11s}. Kerr combs are spontaneously generated from quantum fluctuations in solid state nonlinear optical microresonators pumped with continuous wave light. Since the first demonstration of a comb observed in a fused silica microresonator \cite{delhaye07n}  the possibility of  generating combs in other types of resonators have been reported in numerous experiments. In accordance with experimental observations, the observed combs preferably occur in resonators characterized with anomalous group velocity dispersion (GVD). This conclusion has also been confirmed theoretically \cite{agha07pra,agha09oe}.  Recent numerical simulations conclude that generation of Kerr combs is also possible and stable at all but zero GVD \cite{chembo10pra}.

Generation of the Kerr comb results from  modulation instability (MI) of the continuous wave light propagating in a nonlinear host material of an optical resonator. The physical origin of the instability is the same as those observed in optical fibers, where MI is observed under conditions of anomalous GVD and positive cubic nonlinearity. It is natural to expect that the resonant MI has properties similar to the non-resonant MI. However, it was noted that MI also takes place in a fiber ring resonator with a net normal GVD \cite{haelterman92ol}. The reason is that the resonant configuration introduces an additional degree of freedom in the system, namely frequency detuning of the pumping light from the eigenmode of the nonlinear resonator \cite{coen97prl}; this can shift the MI instability point to the region of normal GVD.

There have been several experiments in which the comb formation was achieved in microresonators with normal GVD. For instance, a comb was observed in a CaF$_2$ whispering gallery mode (WGM) resonator with a diameter of 0.78~cm, pumped with 1320~nm light, and characterized with $\beta_2 \simeq 7.7$~ps$^2/$km normal GVD \cite{savchenkov08oe}. A comb was also generated in a CaF$_2$ WGM resonator featuring a diameter of  0.255~cm, pumped with 1550~nm light, and having $\beta_2 \simeq 2.3$~ps$^2/$km \cite{savchenkov08prl}.

The experimental results were explained theoretically. It was shown that the resonant hyper-parametric oscillation can occur at any GVD (the term "hyper-parametric oscillation" was introduced to describe the underdeveloped Kerr frequency combs, containing only two symmetric sidebands with respect to the carrier) \cite{matsko05pra,matsko11nlo}.

The excitation dynamics represents the basic difference in  Kerr combs generated in resonators with normal and anomalous GVD. Kerr combs can have both hard and soft oscillation onset in the case of anomalous GVD, but only the hard oscillation onset takes place in the case of normal GVD, as noted in \cite{matsko11nlo}. Nevertheless, a more rigorous study of the subject is still required. The theory in \cite{matsko11nlo} was developed for a hyper-parametric oscillator only, and does not cover generation of  Kerr frequency combs. The numerical simulations \cite{chembo10pra} do not describe hard excitation of the comb and do not allow finding any steady state solution for the comb, since it is assumed that the generated sidebands are always smaller compared to the optical pump.

In this Letter we present results of our numerical simulations of Kerr frequency combs produced in a nonlinear microresonator with normal GVD. For the sake of specificity, we present a steady state solution for the comb generated in a CaF$_2$ WGM resonator pumped at 1554~nm. The resonator is characterized with 667~$\mu$m diameter, $100$~GHz FSR, $n_0=1.43$ refractive index, and $\beta_2 \simeq 0.055$~ps$^2/$km GVD, corresponding to 110~kHz frequency difference between adjacent FSRs ($\nu_++\nu_--2\nu_0 \simeq -110$~kHz, where $\nu_0$ is the linear frequency of the pumping light, $\nu_\pm$ are the frequencies of the optical sidebands, $\nu_+-\nu_0\simeq \nu_0-\nu_-\simeq 100$~GHz). The resonator host material has cubic nonlinearity $n_2=1.9\times 10^{-16}$~cm$^2/$W, and mode volume ${\cal V} =1.3\times10^{-7}$~cm$^3$. We assume that the full width at the half maximum of the overloaded WGMs is $2 \gamma_0=220$~kHz, so that $\nu_++\nu_--2\nu_0 = -\gamma_0$.

To find the GVD we used CaF$_2$ Sellmeier equation \cite{daimon02ao} and an asymptotic expression describing the spectrum of a dielectric spherical resonator \cite{lam92josab}. The overall dispersion $\beta_2(\nu_0)$ was calculated using the formula $\nu_++\nu_--2\nu_0 = - \pi c\beta_2 (\nu_0)(\nu_+-\nu_-)^2/2 n_0(\nu_0)$ \cite{matsko03josab}. The selected value of GVD ($\beta_2$) seems to be small, since it is much smaller compared with the dispersion of a conventional optical fiber, however the significance of GVD is determined by the ratio $|\nu_++\nu_--2\nu_0|/\gamma_0$ in a resonator \cite{matsko05pra}. The GVD is considered small if this parameter is much less than unity. Since the difference $\nu_++\nu_--2\nu_0$ depends on the FSR, larger resonators have smaller relative GVD than smaller resonators. Resonators with higher Q-factor have larger relative dispersion compared with lower Q resonators.

To study the behavior of the comb numerically it is convenient to introduce a coupling constant \cite{matsko05pra} $g=\hbar \omega_0^2 c n_2/({\cal V}n_0^2)=3.3\times10^{-3}$~s$^{-1}$, where $\omega_0=2 \pi \nu_0$. Selecting the pump power $P=36.7\ \mu$W, we also introduce the dimensionless pumping constant $f=(F_0/2\pi \gamma_0)(g/2\pi \gamma_0)^{1/2}=2.244$, where $F_0=(4 \pi \gamma_0 P/(\hbar \omega_0))^{1/2}$ describes the amplitude of the continuous wave external pump. The selection of the  above listed parameters is an arbitrary one, simply serving the goal of comparison of the results obtained in our numerical simulations with parameters of a realizable system.

In our model we numerically study the nonlinear interaction of 21 optical modes. The selected number of  modes is limited by the  available computational capacity. We expect that all the modes are identical and completely overlapping in space. The external continuous wave pump is applied to the central mode of the mode group, so the simulated Kerr comb is expected to have ten red- and ten blue-detuned harmonics with respect to the frequency of the pumped mode. We take into account only the second order frequency dispersion that is recalculated to the frequency of the modes \cite{chembo10pra}.

To write the nonlinear equations we introduce an interaction Hamiltonian $\hat V= -g (\hat e^\dag)^2 \hat e^2/2$, where $\hat e= \sum_{j=1}^{21} \hat a_j$. The equations of motion are \cite{chembo10pra,matsko05pra}
\begin{equation} \label{set}
\dot{\hat a}_j=-(\gamma_0+i\omega_j)\hat a_j+ \frac{i}{\hbar} [\hat V,\hat a_j]+F_0 e^{-i\omega t} \delta_{11,j},
\end{equation}
where $\delta_{11,j}$ is the Kronecker's delta. The set of nonlinear forced  equations, (\ref{set}), is solved numerically without any further assumptions. The evaluation was interrupted when the solution reached its steady state ($\gamma_0 t_{s} \approx 30$).
\begin{figure}[htbp]
  \centering
  \includegraphics[width=8.5cm]{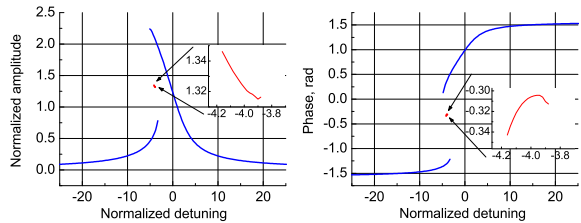}
\caption{ Normalized slow amplitude ($|A_{11}|(g/2\pi\gamma_0)^{1/2}$) and phase of the intracavity field of the pumped mode versus normalized detuning ($(\omega-\omega_{11})/2\pi\gamma_0$). The region of the maximum field accumulation is shifted to smaller frequencies due to the self-phase modulation effect (the effective index of refraction increases with intensity of light circulating in the mode). The pump power is fixed, $f=(F_0/2\pi \gamma_0)(g/2\pi \gamma_0)^{1/2}=2.244$. The stable solutions of the set (\ref{set}) that do not reveal the Kerr comb generation are shown by blue lines. The stable solution that reveals the comb generation is shown by red line. The solution is localized in both  amplitude and frequency spaces. The inset shows the region of parameters where the comb is generated.} \label{fig1}
\end{figure}

\begin{figure}[htbp]
  \centering
  \includegraphics[width=6.5cm]{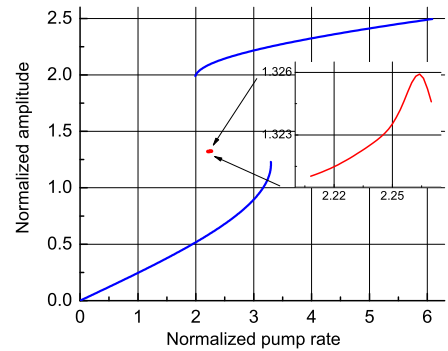}
\caption{ Normalized slow amplitude ($|A_{11}|(g/2\pi\gamma_0)^{1/2}$) and phase of the intracavity field of the pumped mode versus the normalized amplitude of the external pump ($f=(F_0/2\pi \gamma_0)(g/2\pi \gamma_0)^{1/2}$). The frequency detuning is fixed $(\omega-\omega_{11})/2\pi\gamma_0=-4$. Stable solutions of the set (\ref{set}) that do not reveal Kerr comb generation are shown by blue lines. The stable solution that reveals  comb generation is shown by the red line. The solution is localized. The inset shows the region of parameters where the comb is generated.} \label{fig2}
\end{figure}

\begin{figure}[htbp]
  \centering
  \includegraphics[width=7.cm]{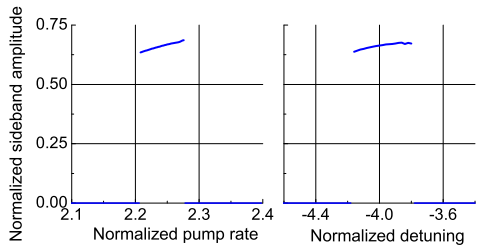}
\caption{ Normalized slow amplitude ($|A_{10}|(g/2\pi\gamma_0)^{1/2}$ and $|A_{12}|(g/2\pi\gamma_0)^{1/2}$ are the same) of the first sidebands of the pump mode  for the same parameters of the system as used in Figs.~(\ref{fig1}) and (\ref{fig2}). } \label{fig3}
\end{figure}

\begin{figure}[htbp]
  \centering
  \includegraphics[width=7.cm]{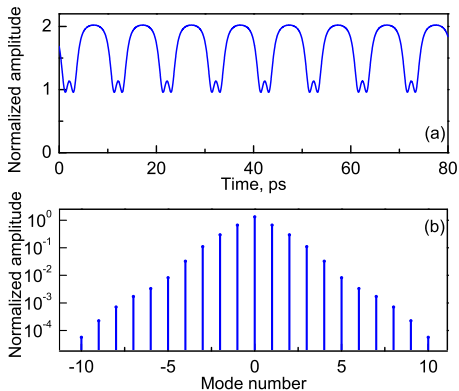}
\caption{Normalized slow amplitude ($(g|\hat e^\dag \hat e|/2\pi\gamma_0)^{1/2}$) of the intracavity field versus time, and the corresponding optical comb found for a fixed pump power $f=2.244$ and frequency detuning of the pump light from the selected mode $(\omega-\omega_{11})/2\pi\gamma_0=-4$. The 100~GHz FSR of the resonator is expected.} \label{fig4}
\end{figure}

The results of simulations are summarized in Figs.~(\ref{fig1})-(\ref{fig4}). To present the results we introduce slow amplitudes of the field in the modes $\hat a_j = A_j \exp(-i\omega_j t)$. The slow amplitudes are further normalized as $|A_{j}|(g/2\pi \gamma_0)^{1/2}$. The frequency detuning is also normalized to the half width at half maximum of the resonator mode, ($(\omega-\omega_{11})/2\pi\gamma_0$). As a rule, the solution followed stable branches (shown by solid blue line) and then jumped to the attractor corresponding to comb generation (shown by the red line). The jump was observed only for the case of nonzero initial conditions. Naturally, the solution does not converge to unstable solutions.

We find that a Kerr comb can be generated in the resonator in the case of normal GVD. For the resonator described above  we obtain the power and the frequency of the pumping light that result in comb excitation. The frequency comb is dynamically stable, and the generation onset is hard in its nature, i.e. there is a discontinuous jump in parameters of the comb harmonics when the pump power exceeds a certain threshold.

The behavior of the field of the mode pumped optically is illustrated by Figs.~(\ref{fig1}) and (\ref{fig2}). If the comb sidebands are small the {\em stable} oscillation occurs in the region that is {\em unstable}. This stability region was not found in previous research \cite{chembo10pra} since only  solutions with adiabatically stable amplitude of the pump mode were analyzed there. The corresponding behavior of amplitudes of the first two optical sidebands ($A_{10}$ and $A_{12}$) is shown in Fig.~(\ref{fig3}). The stability region where the sidebands are generated is localized with respect to the frequency and power of the pumping light.

We found the time dependence of the overall field amplitude in the resonator (Fig.~\ref{fig4}) by selecting specific values of power and frequency of the pump light. Apparently, the modes of the frequency comb create a single optical pulse traveling in the resonator. The shape of the pulse is rather complex and more effort is required to derive an analytical expression for its description. We verified that if the power of the external pump is changed a similar pulse shape, but with different amplitude and duration, is generated in the resonator .

The initial conditions and parameters of equations are varied to verify the stability of the solution. We also reduced and increased the number of modes in our code and confirmed that the spectrum does not change significantly. It is worth noting that the method is suitable for the simulation of a resonator with arbitrary GVD. The  limited number of modes used in the simulation imposes a practical restriction on the values of the pump power and GVD values on the outcome of the simulations.  We intentionally selected a large value of the relative GVD to suppress this undesirable effect.

To conclude, we have studied Kerr frequency comb generation in microresonators made out of a transparent optical material possessing positive cubic nonlinearity, as well as normal group velocity dispersion.  We find that there exists a stable attractor for oscillation that results in  comb generation.  Stable Kerr frequency combs can be produced in such resonators if the proper power and frequency for the continuous wave light pumping a mode of the resonator is selected.

This work was supported in part by DARPA MTO (IMPACT program).

\end{document}